\input{aipcheck}

\documentclass[preprintnumbers
    ,final            
  ]
  {aipproc}

\layoutstyle{8x11single}

\begin{document}

\title{\hspace*{14.0cm}{\small LBNL-59103}\\
The Quark-Gluon Plasma: a Perfect Thermostat and a Perfect Particle Reservoir}

\classification{25.75.-q,25.75.Dw,25.75.Nq,13.85.-t}
\keywords      {Hagedorn mass spectrum, Hagedorn thermostat, statistical hadronization}

\author{L. G. Moretto, K. A. Bugaev, J. B. Elliott and L. Phair}{
  address={Nuclear Science Division, 
Lawrence Berkeley National Laboratory, Berkeley,
California 94720}
}

\begin{abstract}
A system $\cal H$ with a Hagedorn-like mass spectrum imparts its unique temperature $T_{\cal H}$ to any other system coupled to it. An $\cal H$ system radiates particles in preexisting physical and chemical equilibrium. These particles form a saturated vapor at temperature $T_{\cal H}$. This coexistence describes a first order phase transition. An $\cal H$ system is nearly indifferent to fragmentation into smaller $\cal H$ systems. A lower mass cut-off in the spectrum does not significantly alter the general picture. { These properties  of the Hagedorn  thermostats naturally  explain a single value of hadronization temperature 
observed in elementary particle collisions at  high energies  and lead to some experimental predictions.
}
\end{abstract}

\maketitle


\section{Introduction}

A system~$A$ with energy $E$ and degeneracy
\begin{equation}
	\rho_A (E) \propto \exp \left( k_A E \right)
\label{therm-deg}
\end{equation}
while seemingly having a partition function of the form
\begin{equation}
	Z (T) = \int \rho_A (E) \exp\left( -{E}/{T} \right) dE
\label{therm-part}
\end{equation}
for all temperatures $T \le 1 / k_A$ in fact admits only \underline{one} temperature $T=T_A=1/k_A$ and it imparts that temperature to \underline{any} system coupled to it.

The partition function of Eq.~(\ref{therm-part}) implies that an external thermostat $B$ which, by definition has $\rho_B (E) \propto \exp \left( - k_B E \right)$, can impart its temperature $T_B =1/k_B$ to the system $A$. This is not so, as can be seen by considering the generating micro-canonical partition
\begin{eqnarray}
	P (x) & = & \rho_A (E-x) \rho_B (x) =
	\exp \left( k_A \left[ E-x \right] \right) \exp(k_B x) \nonumber \\
	& = & \exp \left[ \frac{E-x}{T_A} \right] \exp \left[ \frac{x}{T_B} \right] .
\label{two-parts}
\end{eqnarray}
The most probable partition is given by
\begin{equation}
	\frac{\partial P(x)}{\partial x} = 0 = k_A - k_B = \frac{1}{T_A} - \frac{1}{T_B} .
\label{mpp}
\end{equation}
But this is hardly possible since in general $T_A \ne T_B$: two thermostats can never be at equilibrium unless they are at the same temperature.

This preamble is motivated by the fact that the empirical hadronic mass spectra (Hagedorn spectra \cite{hagedorn-65,hagedorn-68}), the Statistical Bootstrap Model (SBM) \cite{SBM, SBM:new, SBM:width} and the MIT bag model \cite{chodos-74} have a degeneracy whose leading term is of the form of Eq.~(\ref{therm-deg}). It is the aim of this paper to explore in a pedagogical manner the implications of such a spectrum, making only passing references to the more complex physical situations occurring in particle-particle and nucleus-nucleus collisions. 

Hagedorn noted that the hadronic mass spectrum (level density) has the asymptotic ($m \rightarrow \infty$) form
\begin{equation}
	\rho_{\cal H}(m) \approx \exp \left({m}/{T_{\cal H}}\right)\,,
\label{hagedorn}
\end{equation}	
where $m$ is the mass of the hadron in question and $T_{\cal H}$ is the temperature associated with the mass spectrum \cite{hagedorn-65,hagedorn-68}. The question of the mass range over which (\ref{hagedorn}) is valid is still under discussion \cite{SBM:new,SBM:width}.

The M.I.T. bag model \cite{chodos-74} of partonic matter reproduces this behavior via a constant pressure $B$ of a ``bag'' of partonic matter \cite{Kapusta:81,Kapusta:82}. The pressure $p$ inside a bag at equilibrium without additional conserved quantities is
\begin{equation}
 	p=\frac{g\pi^2}{90} T^{4}_B = B\,,
\label{bagP}
\end{equation}
where $g$ is the number of partonic degrees of freedom. The bag constant forces a constant temperature $T_B$ on the bag. Similarly, the enthalpy density $\epsilon$ of the bag 
\begin{equation}
\label{bagE}
 	\epsilon=\frac{H}{V}=\frac{g\pi^2}{30}T^{4}_B + B\,
\end{equation}
is constant.
Here $H$ is the enthalpy  and $V$ is the volume of the bag. Thus, an injection of an arbitrary amount of energy leads to an isothermal, isobaric expansion of the bag and the bag entropy $S$ is proportional to $H$:
\begin{equation}
	S = \int \frac{\delta Q}{T} = \int_{0}^{H} \frac{dH}{T} = \frac{H}{T_B}\,, 
\label{bag-eq}
\end{equation}
where $\delta Q$ is the change in heat of the bag. The bag's spectrum (level density) is then $\rho = \exp \left( S \right)$ given by Eq.~(\ref{hagedorn}) with $T_{B}=T_{\cal H}$ and $H \equiv m$.

{ Following our recent results \cite{Moretto:05,Bugaev:05}, }
we show here that a system $\cal H$ possessing a Hagedorn-like spectrum, characterized by an entropy of the form (\ref{bag-eq}), not only has a unique microcanonical temperature
\begin{equation}
	T_{\cal H} = \left( \frac{d S}{d E}\right)^{-1} = \left. \frac{\partial H}{\partial S} \right|_p= T_B \,,
\label{temperature}
\end{equation}
but also imparts this same temperature to any other system to which $\cal H$ is coupled. In the language of standard thermodynamics: $\cal H$ is a perfect thermostat.

The property of a perfect thermostat is well known. For instance, it is indifferent to the transfer of any portion of its energy to any parcel within itself, no matter how small. In other words, it is at the limit of phase stability and its internal fluctuations of the energy density are maximal.

\section{Harmonic Oscillator Coupled to $\cal H$}

In order to demonstrate the thermostatic behavior of a Hagedorn system, let us begin by coupling $\cal H$ to a one dimensional harmonic oscillator and use a microcanonical treatment. The unnormalized probability $P(\varepsilon)$ for finding an excitation energy $\varepsilon$ in the harmonic oscillator out of the system's total energy $E$ is
\begin{eqnarray}
	P(\varepsilon) & \sim & \rho_{\cal H}(E-\varepsilon)\, \rho_{\rm osc} (\varepsilon) \nonumber \\
	&=& \exp \left( \frac{E-\varepsilon}{T_{\cal H}} \right) = \rho_{\cal H}(E) \exp \left( -\frac{\varepsilon}{T_{\cal H}} \right).
\label{sho}
\end{eqnarray}
Recall that for a one dimensional harmonic oscillator $\rho_{\rm osc}$ is a constant. The energy spectrum of the oscillator is canonical up to the upper limit ${\varepsilon}_{max} = E$ with an inverse slope (temperature) of $T_{\cal H}$ independent of $E$. The mean value of the energy of the oscillator is given by
\begin{equation}
	\overline{\varepsilon} = T_{\cal H} \left[ 1 - \frac{E / T_{\cal H}}{\exp\left( E / T_{\cal H} \right)-1} \right] .
\label{ave-ho-e}
\end{equation}
Thus in the limit that $E \rightarrow \infty$: $\overline{\varepsilon} \rightarrow T_{\cal H}$, i.e. no temperature other that $T_{\cal H}$ is admitted. { In the standard language of statistical mechanics
this example means that a one dimensional harmonic oscillator  can be used as an ideal thermometer. }

\section{An ideal vapor coupled to $\cal H$}

For a more physically relevant example, let us consider a vapor of $N\gg1$ non-interacting
{ Boltzmann} particles of mass $m_B$ {  and degeneracy $g_B$ } coupled to $\cal H$. The microcanonical level density of the vapor with kinetic energy $\varepsilon$ is
\begin{equation}
 \rho_{\rm vapor}(\varepsilon) = \frac{V^N~g_B^N}{N!\left( \frac{3}{2}N \right)!} 
\left( \frac{m_B \varepsilon}{2\pi} \right)^{\frac{3}{2}N}\,,
\label{vapor-part}
\end{equation}
where $V$ is is the volume. The microcanonical partition of the total system is 
\begin{eqnarray}
	\rho_{\rm total}(E,\varepsilon) & = & \rho_{\cal H}(E-\varepsilon)\rho_{\rm vapor}(\varepsilon) \nonumber \\
& = & \frac{V^N ~g_B^N}{N! \left( \frac{3}{2}N \right)!} \left( \frac{m_B \varepsilon}{2\pi} \right)^{\frac{3}{2}N} e^{\frac{E-m_B \,N-\varepsilon}{T_{\cal H}}} .
\label{full-part}
\end{eqnarray}
Just as with the harmonic oscillator, the distribution of the vapor is exactly canonical up to $\varepsilon_{max}=E$, if the particles are independently present, or $\varepsilon_{max}=E-mN$, if the particles are generated by $\cal H$. In either case, the temperature of the vapor is always $T_{\cal H}$.

The maximum of $\rho_{\rm total}(E,\varepsilon)$ with respect to $\varepsilon$ gives the most probable kinetic energy per particle as
\begin{equation}
	\frac{\partial \rho_{\rm total}(E,\varepsilon)}{\partial \varepsilon} = 
	\frac{3N}{2\varepsilon} - \frac{1}{T_{\cal H}} = 0 \quad \Rightarrow \quad 
	\frac{\varepsilon}{N} = \frac{3}{2}T_{\cal H}\,,
\label{max-01}
\end{equation}
provided that $E \ge m_B\,N + \frac{3}{2}N T_{\cal H}$. (For $m_B\,N < E < m_B\,N + \frac{3}{2}N T_{\cal H}$, the most probable value of the kinetic energy per particle is $\frac{\varepsilon}{N} = \frac{E}{N} - m_B < \frac{3}{2} T_{\cal H}$; for $E \le m_B\,N$, $\frac{\varepsilon}{N}~=~0$.~)\ \ Again $T_{\cal H}$ is the sole temperature characterizing the distribution up to the microcanonical cut-off, which may be above or below the maximum of the distribution since the form of $\rho_{\rm total}(E,\varepsilon)$ is independent of~$E$.

The maximum of $\rho_{\rm total}(E,\varepsilon)$ with respect to $N$ at fixed $V$ is given by
\begin{equation}
	\frac{\partial \ln \rho_{\rm total}(E,\varepsilon)}{\partial N} = -\frac{m_B}{T_{\cal H}} + \ln\left[ g_B \, \frac{V}{N} \left( \frac{m_B\,T_{\cal H}}{2 \pi}\right)^{\frac{3}{2}}\right]= 0,
\label{max-02}
\end{equation}
where Eq.~(\ref{max-01}) was used for $\varepsilon$. Thus the most probable particle density of the vapor is 
independent of $V$:
\begin{equation}
	\frac{N}{V} = g_B\, \left( \frac{m_B\, T_{\cal H}}{2 \pi} \right)^{\frac{3}{2}} e^{-\frac{m_B}{T_{\cal H}}} \equiv n_{\cal H} \,.
\label{numberpp}
\end{equation}
Equation~(\ref{numberpp}) demonstrates that not only is $\cal H$ a perfect thermostat but also a perfect particle reservoir. Particles of different mass $m$ will be in chemical equilibrium with each other. At equilibrium, particles are emitted from $\cal H$ and form a saturated vapor at coexistence with $\cal H$ at temperature $T_{\cal H}$. This describes a first order phase transition (hadronic to partonic). Coexistence occurs at a single temperature fixed by the bag pressure.

These results explain the common value of: the hadronization temperatures obtained within the statistical hadronization model \cite{Becattini:1}; the inverse slopes of the transverse mass spectra of hadrons observed in high energy elementary particle collisions \cite{alexopoulos-02,Gazdzicki:04}; and the transition temperature from lattice QCD calculations for low baryonic density \cite{Lattice}. For further discussion see \cite{Bugaev:05}.

\section{$\cal H$ as a radiant bag}

Let us assume that $\cal H$ is a bag thick enough to absorb any given particle of the vapor striking it. Then, detailed balance requires that on average $\cal H$ radiates back the same particle. Under these conditions particles can be considered to be effectively emitted from the surface of $\cal H$. Thus the relevant fluxes do not depend in any way upon the inner structure of $\cal H$. 

In fact, the results given in equations (\ref{max-01}) and (\ref{numberpp}) show that the saturated vapor concentration depends only upon $m_B$ and $T_{\cal H}$ as long as $\cal H$ is present. A decrease in the volume $V$ does not increase the vapor concentration, but induces a condensation of the corresponding amount of energy out of the vapor and into $\cal H$. An increase in $V$ keeps the vapor concentration constant via evaporation of the corresponding amount of energy out of $\cal H$ and into the vapor. This is reminiscent of liquid-vapor equilibrium at fixed temperature, except that here coexistence occurs at a single temperature $T_{\cal H}$, rather than over a range of temperatures as in ordinary fluids.

The bag wall is Janus faced: one side faces the partonic world, and, aside from conserved charges, radiates a partonic black body radiation responsible for balancing the bag pressure; the other side faces the hadronic world and radiates a hadronic black body radiation, mostly pions. Both sides of the bag wall are at the temperature $T_{\cal H}$. It is tempting to attribute most, if not all, of the hadronic and partonic properties to the wall itself, possibly even the capability to enforce conservation laws globally (quantum number conductivity). Despite the fact that this wall is an insurmountable horizon, with hadronic measurements such as bag size and total radiance we can infer some properties of the partonic world, e.g. the number of degrees of freedom \cite{alexopoulos-02}.

We can estimate an upper limit for the emission time using the outward energy flux of particles radiated from the bag. At equilibrium the in-going and out-going fluxes must be the same, thus the outward flux of particles 
in the nonrelativistic approximation using Eq.~(\ref{numberpp}) is
\begin{equation}
	\varphi_{n_{\cal H}} \simeq \frac{n_{\cal H}}{4} \left( \frac{m_B}{m_B+2T_{\cal H}}\right) \sqrt{8\frac{T_{\cal H}}{\pi m_B}}\,. 
\label{part-flux}
\end{equation}
Using the technique developed in \cite{Bugaev:96, Bugaev:99}, one finds the energy flux $\varphi_{E_{\cal H}}$ and momentum flux 
$p_{\rm rad}$ as
\begin{equation}
	\varphi_{E_{\cal H}} \simeq \left( m_B +2 T_{\cal H}\right) \varphi_{n_{\cal H}} \,, \quad 
	p_{\rm rad} = \frac{1}{2} n_{\cal H} T_{\cal H}\,.
\label{enrgy-flux}
\end{equation}
The pressure $p_{\rm rad}$ exerted on the bag by its radiation can be compared to the intrinsic bag pressure in Eq.~(\ref{bagP}): for pions $p_{\rm rad} \sim 0.02 B$. The time $\tau$ for the bag to dissolve into its own radiation is approximately
\begin{equation}
\tau \simeq \frac{3 \pi ~ \exp 
\left( \frac{m_B}{T_{\cal H}} \right) E_0}{g_B \left( m^2_B \,T_{\cal H}^{2} \right ) R_{0}^{2}}\,,
\label{bag-time}
\end{equation}
where 
$R_0$ is the initial bag radius and $E_0$ is the initial bag total energy.

The fluxes written in Eqs.~(\ref{part-flux}) and (\ref{enrgy-flux}) (particle or energy per unit surface area) 
are integrated over an assumed spherical bag to give the result in Eq.~(\ref{bag-time}). However, because 
of the lack of surface tension, the bag's maximum entropy corresponds to either an elongated (cylinder) or 
a flattened shape (disc). Thus, Eq.~(\ref{bag-time}) should be interpreted as an upper limit. More detailed studies of hadron emission from bags concerning hydrodynamic shock waves and freeze out shocks can be found elsewhere \cite{Bugaev:96, Bugaev:99,studies,studies2}.

The decoupling between the vapor concentration and $m_B$ and $T_{\cal H}$ occurs when $\cal H$ has completely evaporated (i.e. when $E- m_B\,N-\frac{3}{2}NT_{\cal H}=0$) at a volume of 
\begin{equation}
	V_{d} \simeq \frac{1}{n_{\cal H}} \frac{E}{\left[ m_B+\frac{3}{2}T_{\cal H} \right] } .
\label{v-decoup}
\end{equation}
The disappearance of $\cal H$ allows the vapor concentration to decrease inversely proportionally to $V$ as
\begin{equation}
	\frac{N}{V} = \frac{n_{\cal H}V_{d}}{V} .
\label{conc-dec}
\end{equation}

\begin{figure}

  \begin{minipage}[t]{16.cm}
  \hspace*{2.40cm}   \includegraphics[width=10.6cm,height=10.6cm]{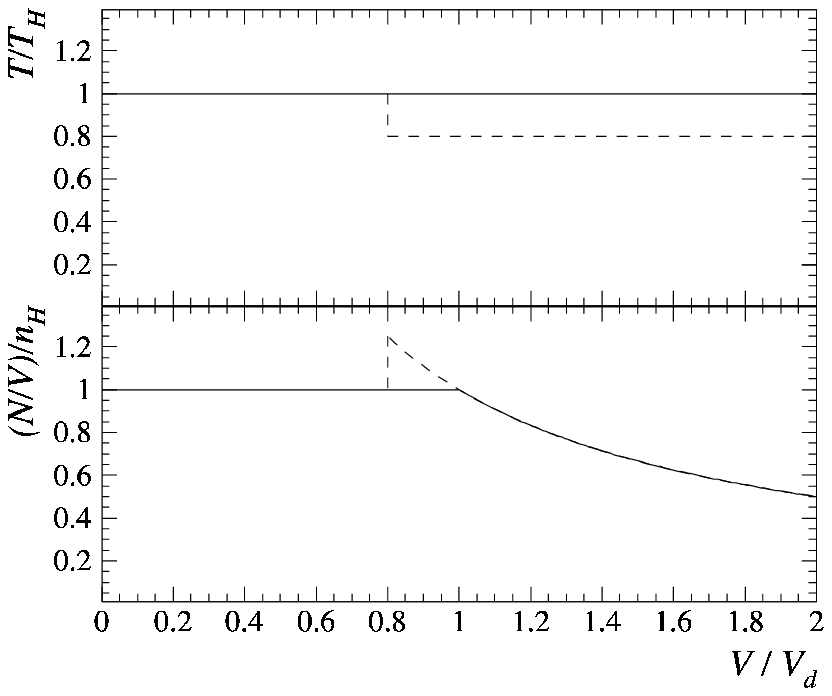}
   
 \noindent  
{{\bf FIGURE 1.}  Typical behavior of the entire system's temperature $T$ and concentration $N/V$ as the  function of  the system's volume $V$ in the absence of restrictions (solid curve) and for a finite cut off at $m_0$  of the Hagedorn spectrum (dashed curve).
}
 \end{minipage}
\end{figure}



The temperature, however, remains fixed at $T_{\cal H}$ because of conservation of energy and particle number above $V_{d}$.\ \ Solid curves in Fig.~1 show this schematically.

The discussion above assumes that the Hagedorn spectrum extends down to $m=0$. However, experimentally there appears to be a lower cut off of the spectrum at $m_0$. This modifies the above results as follows 
{ (for a detailed analysis see the section ``Generalization to a Complete Hagedorn Spectrum'').}

For energies $E - m_B\,N - \varepsilon \gg m_0$ and $V<V_d$ the above results hold as written. However, if we increase the volume well beyond $V_d$ at which the Hagedorn spectrum is truncated at $m_0$, the situation is slightly different. $\cal H$ evaporates until its mass is $m_0$. If the entire mass of $\cal H$ is fully transformed into vapor particles as the volume is increased further, then the excess particles temporarily increase the concentration and permanently decrease the temperature. As the volume increases further, the concentration changes inversely proportional to $V$
\begin{equation}
	\frac{N}{V} = \frac{n_{\cal H} V_{d}+\frac{m_0}{m_B}}{V} ,
\label{conc-dec2}
\end{equation}
\mbox{while the temperature remains constant at}
\begin{equation}
	T = \frac{n_{\cal H} V_d}{n_{\cal H} V_{d}+\frac{m_0}{m_B}} T_{\cal H} .
\label{conc-temp}
\end{equation}
Dashed curves in Fig.~1 show this schematically.

\section{Fragmentation of $\cal H$}

A question of interest is the stability of $\cal H$ against fragmentation. If the translational degrees of freedom are neglected, $\cal H$ is indifferent to fragmentation into an arbitrary number of particles of arbitrary mass (within the constraints of mass/energy conservation).

Let us now consider the case in which the mass of the vapor particle $m_B$ is allowed to be free. The system's level density $\rho_{\rm total}(E,\varepsilon)$ is still given by Eq.~(\ref{full-part}). Using Eqs.~(\ref{max-01}) and~(\ref{numberpp}), one finds the most probable value of the system's level density as $\rho_{\rm total}^*(E,\varepsilon) \approx \exp\left[ S^* \right]$, where the entropy is $S^* = E/T_{\cal H} + N $. Differentiating $\rho_{\rm total}^*(E,\varepsilon)$ with respect to $m_B$ and applying Eq.~(\ref{numberpp}) gives
\begin{equation}
\hspace{-.004cm}\frac{\partial \ln \rho_{\rm total}^*(E,\varepsilon)}{\partial m_B} = N  \left[ \frac{3}{2 m_B}- 
\frac{1}{T_{\cal H} } \right] \,\,\, \Rightarrow \,\,\,  m_B~ = ~\frac{3}{2}T_{\cal H}\,,
\label{masspp}
\end{equation}
i.e. the last equality provides the maximum of level density for $N \neq 0$. Since all the intrinsic statistical weights in $\rho_{\rm total}^*(E,\varepsilon)$ are factored into a single $\cal H$, the system breaks into fragments with $m_B =\frac{3}{2}T_{\cal H}$ except for one whose mass is determined by mass/energy conservation.

Substituting the most probable value of $\varepsilon$ and $m_B$ into the most probable value of $N$ one obtains the vapor concentration
\begin{equation}
	\frac{N}{V} =  g_B \left( \frac{3}{4 \pi e} \right)^{\frac{3}{2}} T_{\cal H}^{3}\,.
\label{concpp}
\end{equation}
The density of the vapor of nonrelativistic particles acquires the form typical of the ultrarelativistic limit.

If the value of mass given by Eq.~(\ref{masspp}) does not exist, then the most probable
value of level density $\rho_{\rm total}^*(E,\varepsilon) $ corresponds to the mass $m^*$ which is nearest to 
$\frac{3}{2}T_{\cal H}$ 
and $N (m^*) $ given by Eq. (\ref{numberpp}). 
In terms of hadron spectroscopy the value of $m^*$ that maximizes  
the level density $\rho_{\rm total}^*(E,\varepsilon) $
is the pion mass. 

If $\cal H$ is required to fragment totally into a number of equal fragments { of mass $m_H$} all endowed with their translational degrees of freedom, then { (for  $g_B = 1$) }
\begin{eqnarray}
\rho_T & = &
 \frac{ {e^{\frac{E - \varepsilon}{ T_{\cal H} } } }
 V^N}{N! \left( \frac{3}{2}N\right)!} 
 \left[ \frac{m_H\, \varepsilon}{2 \pi} \right]^{\frac{3}{2}N} 
  = 
 \frac{ {e^{\frac{E}{ T_{\cal H} } } } ~ V^N}{N! } \left[ \frac{m_H\, T_{\cal H}}{2 \pi} \right]^{\frac{3}{2}N} ,
\label{equation}
\end{eqnarray}
where in the last step we substituted the most probable value of the kinetic energy (\ref{max-01}) 
and used the Stirling formula for $ \left( \frac{3}{2}N\right)!$. From 
Eq.~(\ref{equation}) it is seen that all the Hagedorn factors collapse into a single one with the 
$m$-independent argument $E$. Maximization of (\ref{equation}) with respect to $m_H$ leads to
\begin{equation}
	\frac{\partial \ln \rho_T}{\partial m_H} = \frac{3\,N}{2\,m_H} = 0\,,
\label{max-eq}
\end{equation}
which is consistent with $N=1$ and $m_H=E$, namely a single Hagedorn particle with all the available mass. 

This again illustrates the indifference of $\cal H$ toward fragmentation. Of course Eq.~(\ref{max-01}) gives directly the mass distribution of the Hagedorn fragments under the two conditions discussed above. These results justify the assumption of the canonical formulation of the statistical hadronization model that smaller clusters appear from a single large cluster \cite{Becattini:Can}.

\section{Intermediate Conclusions}

A system $\cal H$, with a Hagedorn-like mass spectrum, is a perfect thermostat and a perfect particle reservoir. Consequently, any system coupled to $\cal H$ can have only the temperature of $\cal H$: $T_{\cal H}$. This behavior may explain the common value of: the hadronization temperatures obtained within statistical models; the transition temperature from lattice QCD calculations for low baryonic density; and the inverse slopes of the transverse mass spectra of hadrons (temperature) observed in high energy elementary particle collisions and high energy nucleus-nucleus collisions  (for details see \cite{Bugaev:05}). The common temperature of the experimental spectra suggest that the observed particles originate from an $\cal H$-like system.

The hadronic side of $\cal H$ radiates particles in preexisting physical and chemical equilibrium just as a black body radiates photons in physical and chemical equilibirum (compare to Ref. \cite{Greiner:04}). Particles emitted from $\cal H$ form a saturated vapor that coexists with $\cal H$. This coexistence describes a first order phase transition (hadronic to partonic) and occurs at a single temperature fixed by the bag pressure. An $\cal H$ system is nearly indifferent to fragmentation into smaller $\cal H$ systems. A lower cut-off in the mass spectrum does not alter our results \cite{Bugaev:05}.

\section{Generalization to a Complete Hagedorn Spectrum}

 \vspace*{-0.1cm}

{ To have  a more realistic model we should  consider 
a more complicated  Hagedorn mass spectrum 
$g_H (m_H) = \exp[ m_H/ T_{\cal H} ] (m_{\rm o}/ m_H )^a$ 
for the  resonance masses $m_H$   above the lower  cut-off 
$m_{\rm o} \gg T_{\cal H}$ ($a$ is a parameter discussed below).
Let us study }
the microcanonical ensemble of $N_B$ Boltzmann point-like particles 
of mass $m_B$ and degeneracy $g_B$, 
and $N_H$  hadronic  point-like resonances of mass $m_H $ with a 
mass spectrum $g_H (m_H) $ {  assuming that }
$m_{\rm o} > m_B$.
A recent analysis \cite{Bron:04}  suggests  that the  Hagedorn mass spectrum  can be established  for $m_{\rm o}  < 2$ GeV.

In the Statistical Bootstrap Model (SBM) \cite{Frautschi:71} and the  MIT bag model  \cite{Kapusta:81}
 it was found that for $m_H \rightarrow \infty$  the parameter $a \le 3$.
For finite resonance masses  the  value of $a$  is  unknown, so   it will be   considered as a fixed parameter.

The microcanonical partition 
of the system,  with  volume $V$,  total energy $E$ and zero total momentum,  can be written as follows 

\vspace*{-0.3cm}
\begin{equation}\label{mone}
\Omega
= 
\frac{V^{N_H} }{ N_H !}  \left[ \prod_{k = 1}^{N_H}  g_H(m_H)\hspace*{-0.0cm} \int \hspace*{-0.0cm} 
\frac{ d^3 Q_k}{(2 \pi)^3 } \right] 
%
\frac{V^{N_B} }{ N_B !} ~ \left[ \prod_{l = 1}^{N_B} g_B \hspace*{-0.0cm}\int  \hspace*{-0.0cm}
\frac{ d^3 p_l}{(2 \pi)^3 } \right] ~
\delta\biggl( E - \sum\limits_{i = 1 }^{N_H} \epsilon^H_i - \sum\limits_{j = 1 }^{N_B} \epsilon^B_j \biggl) ,
\end{equation}
where the quantity $ \epsilon^H_i~=~\varepsilon(m_H, Q_i )$ $\left( \epsilon^B_j = \varepsilon(m_B, p_j) \right.$ and
$\left. \varepsilon(M, P) \equiv  \sqrt{M^2 + P^2}  \right)$  denotes  the energy  of
the Hagedorn (Boltzmann) particle with the 3-momentum ${\vec Q}_i$ (${\vec p}_j$).
In order to simplify the  presentation of our idea,  Eq. (\ref{mone}) accounts  for   energy conservation only 
and neglects  momentum conservation.

The microcanonical partition (\ref{mone}) can be evaluated by the Laplace transform 
in total energy $E$ \cite{Pathria}.
Then the momentum integrals in (\ref{mone}) are factorized and can be performed
analytically. The inverse Laplace transform in the conjugate variable $\lambda$ can be 
 done analytically for
 the nonrelativistic and ultrarelativistic approximations  of 
the one-particle momentum distribution  function
\begin{eqnarray}\label{mfour}
\hspace*{-0.cm} 
 \int\limits_0^\infty \hspace*{-0.1cm}  
\frac{d^3 p ~  {\textstyle e^{-\lambda \varepsilon(M , p ) } }   }{ ( 2 \pi)^3 } 
 \approx  
 \left\{
\begin{tabular}{ll}
\vspace{0.1cm} \hspace*{-0.1cm}$ \left[ \frac{ 2 M }{\lambda} \right]^{\frac{3}{2} }\hspace*{-0.1cm} I_{\frac{1}{2} }  e^{- M \lambda } \,,$
&  $  M  Re (\lambda)  \gg 1$\,, \\
\hspace*{-0.1cm}$ \frac{ 2 }{  \lambda^3 } ~ I_{ 2 }\, e^{- M \lambda }  \,, $  & $  M  Re (\lambda)  \ll 1$\,,
\end{tabular}
\right.
\end{eqnarray}
where the auxiliary integral is denoted as 
\begin{equation}\label{mfive}
I_b ~\equiv ~ \int\limits_0^\infty \hspace*{-0.0cm}  
\frac{d \xi }{ (2 \pi)^2 } ~ \xi^b
~{ e^{-\xi } }  \,.
\end{equation}
\vspace*{-0.30cm}

Since the formal steps of further evaluation  
are similar for  both cases, we discuss 
in detail the nonrelativistic limit only,  and later 
present  the results for the other case.  
The  nonrelativistic approximation ($  M  Re (\lambda)  \gg 1$) for  Eq. ~(\ref{mone}) is as follows
\vspace*{-0.1cm}
\begin{eqnarray}
&&
\hspace*{-0.3cm} 
\Omega_{nr} 
= 
\frac{ \left[  V g_H(m_H)  \left[ 2 m_H  \right]^{ \frac{3}{2} }    I_{ \frac{1}{2} }  
\right]}{N_H!}^{N_H}
 \frac{ \left[  V g_B \left[ 2 m_B \right]^{ \frac{3}{2} }  ~ I_{ \frac{1}{2} } \right] }{N_B!}^{N_B} 
\hspace*{-0.2cm} 
\frac{ E_{kin}^{\frac{3}{2} (N_H + N_B) - 1}  }{  \left( \frac{3}{2} (N_H + N_B) - 1 \right)! }\,,
\end{eqnarray}\label{meight}
where $E_{kin} = E - m_H N_H - m_B N_B $ is the kinetic energy of the system.

As shown below, the most realistic case 
corresponds to the nonrelativistic treatment of the Hagedorn resonances because 
the resulting temperature is  much smaller than 
their masses. Therefore, it is sufficient to consider the ultrarelativistic
limit for the Boltzmann particles only. In this case 
 ($  M  Re (\lambda)  \ll 1$)
 the equation (\ref{mone}) can be approximated as 
\begin{eqnarray}\label{mten}
&&\hspace*{-0.6cm} \Omega_{ur} 
= 
\frac{ \left[  V g_H(m_H)  \left[ 2 m_H  \right]^{ \frac{3}{2} }   ~ I_{ \frac{1}{2} }  
\right]}{N_H!}^{N_H}   
 \frac{ \left[  V g_B  ~2  ~ I_{ 2 } \right] }{N_B!}^{N_B} 
\hspace*{-0.0cm} 
\frac{ E_{kin}^{\frac{3}{2} (N_H + 2 N_B) - 1}  }{  \left( \frac{3}{2} (N_H + 2 N_B) - 1 \right)! }\,,
\end{eqnarray}
where the  kinetic energy does not include the rest energy  of the Boltzmann particles, i.e.
$E_{kin} = E - m_H N_H $.

Within our assumptions 
 the above results are general
and can be used for any number of particles,  provided $N_H + N_B \ge 2$.
It is instructive to consider first  the  simplest case $N_H = 1$. 
This formulation of the  model, in which a Hagedorn thermostat is always present,  allows
us to study the problem rigorously and provides us with a qualitative picture for $N_H > 1$.
For  $N_H = 1$ and $N_B~\gg~1$ we treat  the mass of Hagedorn
thermostat $m_H$  as a free parameter and determine the  value which maximizes the entropy of the system.  
The  solution $ m_H^*~>~0$  of 
 the  extremum condition 
\begin{eqnarray}\label{meleven}
\hspace*{-0.0cm}&& \frac{ \delta \ln \Omega_{nr} (N_H = 1) }{\delta~ m_H } ~ 
 {\textstyle  \frac{1}{T_{\cal H}}~ +~ \left( \frac{3}{2}~ - ~ a \right) \frac{1}{m_H^*} ~ - ~  
\frac{3 (N_B + 1) }{2~ E_{kin} } ~ = ~ 0 }  
\end{eqnarray}
provides the maximum of the system's entropy, if  for $m_H = m_H^*$ the second derivative is negative
\begin{eqnarray}\label{mtwelve}
\hspace*{-0.0cm}&& \frac{ \delta^2 \ln \Omega_{nr} (N_H = 1) }{\delta~ m_H^2 } ~ 
{\textstyle - ~ \left( \frac{3}{2}~ - ~ a \right) \frac{1}{m_H^{*\,2} } ~ - ~  \frac{3 (N_B + 1) }{2~ E_{kin}^2 } ~ < ~ 0 \,. }
\end{eqnarray}
 The inequality  (\ref{mtwelve}) is  a necessary condition of the maximum of the 
microcanonical partition.  Postponing  the analysis of  (\ref{mtwelve}) till the next section,  where  we study it in more details,  let us assume for  a  moment  that
the inequality  (\ref{mtwelve}) is satisfied.  Then the extremum condition (\ref{meleven})
defines the  temperature of the system of $(N_B + 1)$ nonrelativistic particles
\begin{equation}\label{mthirteen}
\hspace*{-0.25cm} T^* (m_H^*)  \equiv  
\frac{ 2 ~ E_{kin} }{  3 (N_B + 1) } = \frac{T_{\cal H}}{ 1 ~ + ~ \left(  \frac{3}{2}~ - ~ a \right) \frac{T_{\cal H} }{m_H^{*} }    }  \,.
\end{equation}
Thus, as $m_H^* \rightarrow \infty$ it follows that $T^*(m_H^*) \rightarrow T_{\cal H}$, while for 
finite $m_H^* \gg T_{\cal H}$  and $ a >  \frac{3}{2} $  ($ a <  \frac{3}{2} $) the temperature 
of the system is 
slightly larger  (smaller) than the Hagedorn temperature, i.e. $ T^* > T_{\cal H}$ ($ T^* < T_{\cal H}$).
Formally, the temperature of the system in equation (\ref{mthirteen}) may differ  essentially 
from $T_{\cal H}$ for  a  light thermostat, i.e. for $m_H^* \le T_{\cal H}$.
However, it is assumed that 
the Hagedorn mass spectrum exists
above the cut-off mass $m_{\rm o} \gg T_{\cal H}$, thus $m^* \gg T_{\cal H}$.


\vspace*{-0.3cm}

\section{The Role of the Mass Cut-off}

\vspace*{-0.3cm}

Now we study the effect of the mass cut-off of the Hagedorn spectrum on the
inequality (\ref{mtwelve}) in more detail. 
For $ a \le  \frac{3}{2} $ the condition (\ref{mtwelve}) is  satisfied. For $ a >   \frac{3}{2} $ the inequality (\ref{mtwelve})
is equivalent to the following inequality 
\begin{equation}\label{mfourteen}
\hspace*{-0.25cm}
 \frac{ m_H^{*\,2}  }{   \left( a -  \frac{3}{2} \right)  ~ T^*(m_H^*)    }  ~ > ~ 
\frac{ 3 }{ 2 } ~  (N_B + 1) ~ T^*(m_H^*)  
\,,
\end{equation}
which  means that a  Hagedorn thermostat should be massive  compared to the kinetic energy of the system.  

A more careful analysis shows that 
for a negative value of the determinant $ D_{nr} $  $( \tilde{N} \equiv N_B - \frac{2}{3} a )$  
\begin{eqnarray}\label{mfifteen}
\hspace*{-0.25cm}
D_{nr} & \equiv & {\textstyle  \left( E - m_B N_B - \frac{3}{2}~ T_{\cal H} ~ \tilde{N} \right)^2 - } \nonumber \\ 
& & {\textstyle 4 \left( a - \frac{3}{2} \right)~T_{\cal H}~ \left( E - m_B N_B \right) ~ < ~ 0\,, }
\end{eqnarray}
equation (\ref{meleven}) has two complex solutions, while  for $D_{nr}~=~ 0$ there exists a single 
real solution of (\ref{meleven}).
Solving (\ref{mfifteen}) for $(E - m_B N_B)$, shows that  for $\tilde{N} > \frac{2}{3} a - 1$,
 i.e. for ${N_B} > \frac{4}{3},  a - 1$ the inequality (\ref{mfifteen})  does not hold and $D_{nr} > 0$.
Therefore, in what follows we will assume that ${N_B} > \frac{4}{3} a - 1$ and  only analyze  the case $D_{nr} > 0$. 
For this case  equation (\ref{meleven}) has two real solutions 
\begin{equation}\label{msixteen}
m_H^\pm = {\textstyle \frac{1}{2} \left[ E - m_B N_B - \frac{3}{2}~ T_{\cal H} ~ \tilde{N}~ \pm ~ \sqrt{ D_{nr} } \right]\,.} 
\end{equation}
For $ a \le  \frac{3}{2} $ only $m_H^+$ solution is positive and
corresponds to a maximum of  the microcanonical partition $\Omega_{nr}$. 

For $ a >  \frac{3}{2} $ both solutions of (\ref{meleven}) are positive, but only  $m_H^+$  is a maximum.
From the two limiting cases:
\begin{eqnarray}\label{mseventeen}
\hspace*{-0.5cm}
\frac{ \delta \ln \Omega_{nr} (N_H = 1) }{\delta~ m_H } & \approx &  
{\textstyle \left( \frac{3}{2} -  a \right) \frac{1}{m_H}  \quad {\rm for} \quad m_H \approx 0\,,} 
\\
\label{meighteen}
\hspace*{-0.5cm}
%
\frac{ \delta \ln  \Omega_{nr} (N_H = 1)  }{\delta~ m_H }  & \approx &
{\textstyle 
\frac{3 (N_B + 1) }{2~ E_{kin} } \quad {\rm for} \quad E_{kin} \approx 0\,,}
\end{eqnarray}
and the fact that $ m_H^\pm $ 
obey the   inequalities
\begin{equation}\label{mnineteen}
0 ~ < ~  m_H^- ~  \le ~ m_H^+ ~ < ~ E - m_B N_B \,, 
\end{equation}
it is clear that $ m_H^* = m_H^-$ is a local minimum  of the microcanonical partition $\Omega_{nr}$,
while  $ m_H^* = m_H^+$ is  a local maximum of the partition $\Omega_{nr}$.

Using Eq. (\ref{msixteen}) for $m_H^+$, it is clear that 
for any value of  $a$ 
the constraint $m_H^+ \ge m_{\rm o} $
is equivalent to the  inequality 
\begin{equation}\label{mtwenty}
N_B ~ \le ~ N_B^{kin} \equiv  { \frac{ E ~ - ~ [ \frac{m_{\rm o} }{T_{\cal H}}~ - ~a ] ~ T^*(m_{\rm o})  }{ m_B \, + \, \frac{3}{2} ~ T^*(m_{\rm o})  }  } \,.
\end{equation}
Thus, at fixed energy $E$  for all  $N_B \le N_B^{kin}$ 
 at  $m_H^*  = m_H^+$ there is a local maximum of the microcanonical partition  $\Omega_{nr}$ with
the temperature $ T = T^*( m_H^+) $. For $N_B > N_B^{kin}$   the maximum of the partition $\Omega_{nr}$ cannot be reached due to the cut-off constraint
and, consequently,   the most probable state corresponds to $m_H = m_{\rm o}$
with 
$T  \le T^*(m_{\rm o} )$ from Eq. (\ref{mthirteen}). 
In other words, for $N_B > N_B^{kin}$  the amount of energy $E$ is insufficient for  the mass of the Hagedorn thermostat to be above the cut-off  $m_{\rm o} $ 
and   simultaneously  maintain  the  temperature of the Boltzmann particles according to  
Eq. (\ref{mthirteen}).  
By assumption there is a single Hagedorn thermostat in the system, therefore,  
as $N_B$ grows the temperature of the system decreases
from $T^*(m_{\rm o} )$ value.
Thus, the equality (\ref{mtwenty}) defines the kinematical limit 
for reaching the maximum of the microcanonical partition.

To prove that the maximum of the microcanonical 
partition at $m_H = m_H^+$  is  
global  it is sufficient to show that 
the  constraint $m_H^+ \ge m_{\rm o}$
is not consistent with the condition $m_H^- > m_{\rm o}$. 
For $a \le \frac{3}{2}$ the maximum is  global  because for $ 0 <  m_H < m_H^+$ 
($m_H >  m_H^+$ ) the partition $\Omega_{nr} (N_H =1, m_H ) $ monotonically increases (decreases)
with $m_H$. 
For $a > \frac{3}{2}$  it is clear that the maximum at $m_H = m_H^+$ is local, 
if  the state with mass  $m_H =  m_{\rm o}$ is more probable, i.e. 
$\Omega_{nr} (N_H =1, m_{\rm o}) >  \Omega_{nr} (N_H =1, m_H^+ ) $.  Due to (\ref{mnineteen})  this can occur, if $m_H^- > m_{\rm o}$.  Substituting Eq.  (\ref{msixteen})  into the last inequality,
shows  that this inequality
reduces  to the condition $N_B > N_B^{kin}$.
This contradicts  the  constraint $m_H^+ \ge m_{\rm o}$ in the form of  Eq. 
(\ref{mtwenty}). 
Thus, the maximum of the
microcanonical partition is  global.

To complete our consideration of the  nonrelativistic case 
let us express the partition (\ref{meight}) in terms
of the temperature (\ref{mthirteen}).  Applying the Stirling approximation  to  the  factorial $(\frac{3}{2}(N_B + 1) -1 )!$
for 
$N_B^{kin} > N_B \gg 1$ and reversing the
integral representations (\ref{mfour})  and (\ref{mfive}) for $\lambda = 1/ T^*(m_H^+)$, one
finds 
\begin{eqnarray}\label{mtwone}
&&\hspace*{-0.6cm} \Omega_{nr} (N_H=1) ~ = ~  
%
%
\frac{  V \, g_H(m_H^+)  }{  T^*(m_H^+) } ~  \hspace*{-0.0cm}   \left[ \int  \hspace*{-0.0cm}  
\frac{d^3 Q}{ (2 \pi)^3 } 
~{\textstyle e^{ - \frac{  \sqrt{m^{+\,2}_H + Q^2}   }{  T^*(m_H^+) }  } } \right]  ~
 \frac{ e^{\frac{E}{ T^*(m_H^+) } }  }{N_B!} ~   \left[ V \, g_B  \hspace*{-0.0cm}  \int  \hspace*{-0.0cm}  
\frac{d^3 p}{ (2 \pi)^3 } 
~{\textstyle e^{ - \frac{  \sqrt{m^2_B + p^2}   }{  T^*(m_H^+) }  } }  \right]^{N_B}  \,.
\end{eqnarray}
This is just 
the  grand canonical partition of $(N_B + 1)$ Boltzmann
particles with temperature $ T^*(m_H^+) $. 
If $N_B > N_B^{kin} \gg 1$, then $ T^*(m_H^+)$
in (\ref{mtwone}) should be 
replaced  by 
$T_{\rm o} (N_B) \equiv  \frac{2(E - m_B N_B - m_{\rm o} ) }{3 (N_B + 1) } $. 

Fig. 1 shows that 
for  $a >  \frac{3}{2}  $
the system's temperature $T = T^*(m_H^+)$  as a function of $N_B$ remains almost  constant    for 
 $N_B < N_B^{kin}$, reaches a maximum  at  $N_B^{kin} $  and rapidly decreases like 
 $T  = T_{\rm o}  (N_B) $  for  $N_B > N_B^{kin} $.
For $a <  \frac{3}{2} $ the temperature  has a plateau $T =  T^*(m_H^+)$ for $N_B <  N_B^{kin}$,
and rapidly    decreases  for 
$N_B > N_B^{kin} $ according to  $T_{\rm o}  (N_B) $. 

The same results  are  valid for the ultrarelativistic treatment of Boltzmann 
particles.  Comparing the nonrelativistic and ultrarelativistics expressions for
the microcanonical partition, i.e. equations (\ref{meight}) and (\ref{mten}), respectively,
one  finds that   the derivation of the  ultrarelativistic limit   requires only 
the substitution $N_B \rightarrow 2 N_B$ and $m_B / T_{\cal H} \rightarrow 0$
in  equations (\ref{meleven} -- \ref{mtwone}). 
Note that  this  substitution does not alter  the expression for the temperature
of the system, i.e. the right hand side of (\ref{mthirteen}).

Finally,  we show that for a heavy Hagedorn thermostat  ($m_H^+  \gg  m_{\rm o}$) these results remain 
valid for  a single Hagedorn thermostat split  into $N_H$ pieces of the same mass.
Substituting $m_H \rightarrow m_H N_H$ in the nonrelativistic expressions 
(\ref{meight}) and minimizing  it with respect to $m_H$,  the temperature of the system in the form of  equation (\ref{mthirteen})  is $ T^*(m_H^* N_H) $, where the mass of $N_H$ Hagedorn thermostats $m_H^*$ is related to the solution $m_H^+ $ of equation (\ref{msixteen})   as  
$m_H^* = m_H^+ / N_H$. Since the original single thermostat of mass $m_H^+ $ was assumed to be heavy,  it follows $T^*(m_H^* N_H) = T^*(m_H^+) \rightarrow T_{\cal H}$. 
A more careful study (see also   \cite{Moretto:05})  using
an exact expression for the microcanonical partition of $N_H$ Hagedorn thermostats of the same mass
$m_H$ gives the same result, if  $m_H  \gg  m_{\rm o} $. 
A generalization of these statements to  the case of $N_H$ heavy 
Hagedorn thermostats of different masses  also leads to the same result.  
Thus,  splitting  a single
heavy Hagedorn thermostat into an arbitrary number of heavy resonances (heavier than $ m_{\rm o} $) does not
change the temperature of the system.

%
%
\vspace*{-0.5cm}

\begin{figure}

  \begin{minipage}[t]{16.cm}
  \hspace*{+2.4cm}   \includegraphics[width=10.6cm,height=10.6cm]{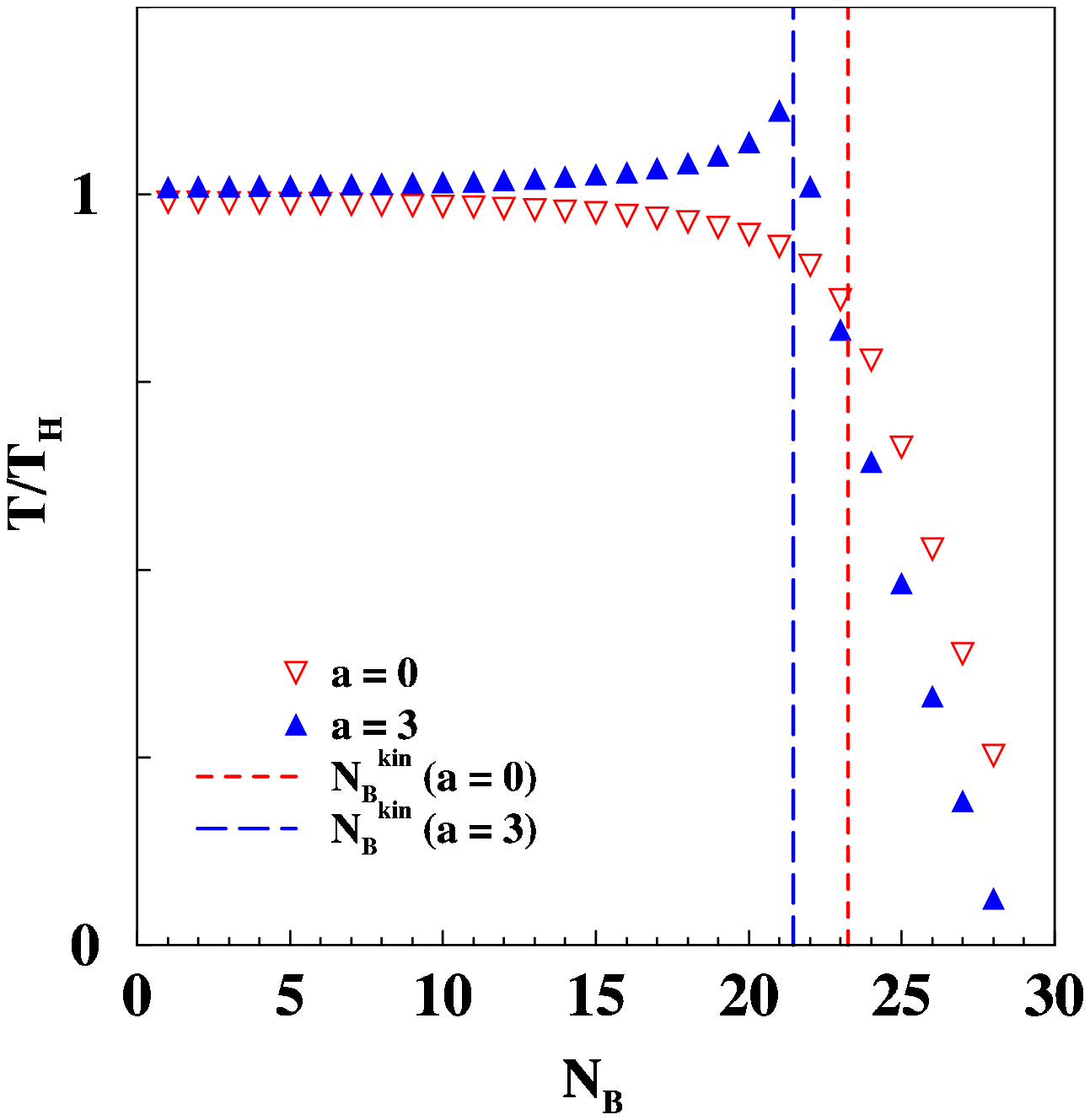}
   
 \noindent  
{{\bf FIGURE 2.} A typical behavior  of the system's temperature  as the function of the  number 
of Boltzmann particles $N_B$ for $ a = 3$ and $a = 0$ for the same value of the total energy 
$E = 30 \,m_B$.
Due to the thermostatic properties of a Hagedorn resonance
the system's temperature  is nearly constant 
up to the kinematically allowed value $N_B^{kin}$ given by (\ref{mtwenty}). 
}
 \end{minipage}
\label{fig2}
\end{figure}



\vspace*{-0.0cm}


\section{The Bag Surface}

\vspace*{-0.2cm}

The bag expressions reported  above contain only volume terms.
Given the finite  size  of the bags that are  typically considered (resonances), it may be of interest to
consider finite size effects and their role in the description of the bags properties. 
The simplest generalization,  assuming  that the bags are leptodermous 
(which is supported by the short range of hadron-hadron interaction and by the saturating properties 
implicit in Eqs. 
 (\ref{bagP}) and (\ref{bagE})), 
 is the  introduction of  surface energy. 
 This can be done  phenomenologically by introducing a $V^{ \frac{2}{3} }$ term in the free energy.
Then   the pressure of  a spherical  bag can be written as 
\begin{equation} \label{Nsix}
p =  \frac{\sigma}{3}  T^4 ~-~ B~ - ~ \frac{2}{3}  a_s(T) \, V^{ - \frac{1}{3} }   = 
 \frac{\sigma}{3}  T^4~ -~ B~ - ~ \frac{2}{3} \frac{a_s(T) }{ \alpha \, R  } \,,
\end{equation}
where $a_s(T)$ is the  temperature dependent surface energy coefficient, $R$ is the bag radius
and  $\alpha \equiv \left[  \frac{ 4 \pi }{3}  \right]^{ \frac{1}{3} }  $. 
Using the thermodynamic identities for the free energy $F$ and entropy $S$
\begin{equation} \label{Nseven}
p = - \left( \frac{ \partial  F}{  \partial V} \right)_T \,, \quad {\rm and} \quad 
S = - \left( \frac{ \partial  F}{  \partial T} \right)_V \,,
\end{equation}
one can find all thermodynamic functions as follows 
\begin{eqnarray}\label{Neight}
\hspace*{-0.cm} 
F  & = &   - \left[  \frac{\sigma}{3}  T^4 ~-~  B \right] V~ + ~   a_s(T) ~V^{ \frac{2}{3} } \,, \\
\label{Nnine}
S  & = &    \frac{4\, \sigma}{3}   T^3 V~ - ~  \frac{d  a_s(T)}{d~~ T}  ~ V^{ \frac{2}{3} }  \,, \\
\label{Nten}
E ~\equiv ~\varepsilon V  &  = &  \left[ \sigma T^4 ~+~ B \right]  V ~ + ~   \left[  a_s(T)~ - ~   \frac{d  a_s(T)}{d~~ T} 
\right]  ~ V^{ \frac{2}{3} }  \,.~
\end{eqnarray}
In evaluating the expression (\ref{Neight})  we fixed the integration constant (an arbitrary function 
of $T$) to zero because  the free energy should vanish
for the bag of zero volume.

While  the magnitude of $ a_s(T) $ is unknown, there  are surprising  consequences  for  $ a_s(T) > 0$.
In Eq. (\ref{Nsix}) the surface term appears as an 
\underline{additional} pressure to the bag pressure.  
Therefore, for a bag in a vacuum  the total pressure should be zero, i.e.
$p = 0$, and, consequently,  the bag temperature acquires  volume dependence: 

\begin{equation} \label{Neleven}
T(R) =   \left[  \frac{3}{\sigma}  \left( B + \frac{2\,a_s(T) }{3\, \alpha \, R  }   \right) \right]^{ \frac{1}{4} }  \,.
\end{equation}
When $R$ is large we recover the previous bag temperature and the associated  physics.
When $R$ becomes small,  however, the bag temperature increases!
The implications of this dependence are strange indeed.
The first is the peculiar behavior of the bag's heat capacity.
The second is the stability of the gas of bags (or lack thereof).
The third is  the signature of  a bag's decay.

\begin{figure}

  \begin{minipage}[t]{16.cm}
   \vspace{-0.5cm}
  \hspace*{-0.25cm}   \includegraphics[width=16.6cm,height=16.0cm]{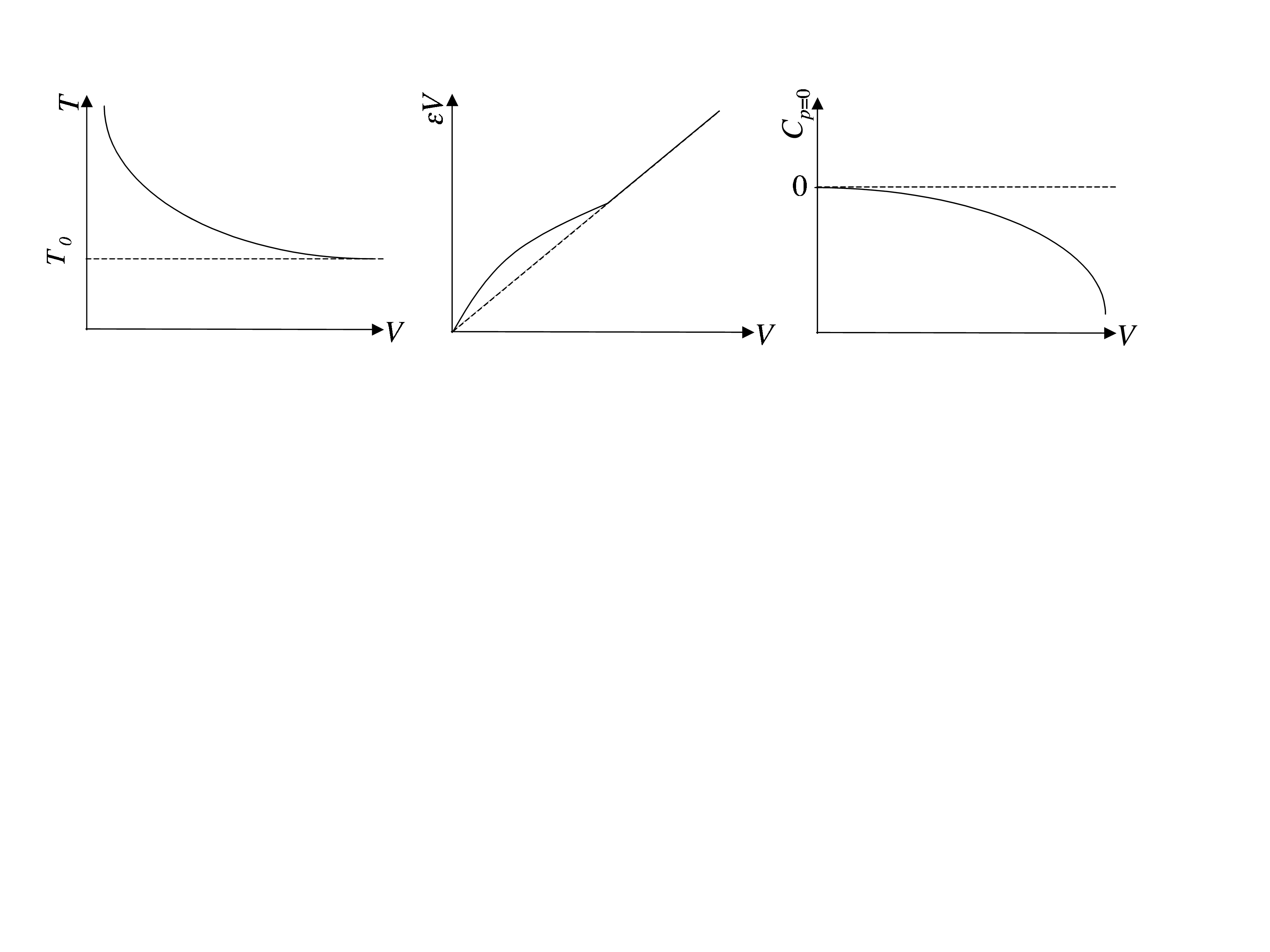}
   \vspace*{-9.5cm}
   
 \noindent  
{{\bf FIGURE 3.}  A schematic volume dependence of
the bag temperature (left panel),  bag energy (middle panel)
and its heat capacity (right panel) for  the  temperature independent surface tension 
$a_s (T) = a_{\rm o} > 0.$ 
The left  and right  panels  show the volume dependence of  the right hand side 
of Eqs. (\ref{Neleven}) and (\ref{Nten}), respectively.  The resulting heat capacity  of the bag
is negative (right panel).  The parameter $T_{\rm o}  $ is defined by the bag constant 
as follows: $ T_{\rm o} = \left[ 3\, B/ \sigma \right]^{1/4} . $
}
 \end{minipage}
\label{fig3}
\end{figure}


\vspace*{-0.3cm}

\section{Heat Capacity}

\vspace*{-0.2cm}

In the standard bag model the heat capacity is infinite: no matter how much energy is fed 
to the bag, its temperature remains constant \cite{Moretto:05,Bugaev:05}.  The only effect is to make the bag larger.  This is completely consistent with what we observe in isobaric  phase transitions 
in ordinary matter.  Here the isobaric condition is produced by the bag constant, and the phase
transition is from hadronic to partonic phase. 

Including surface effects,  shows that 
the more  energy is  put into the bag, the lower its temperature becomes: i.e.
the bag's heat capacity is negative.
To illustrate how the negative heat capacity of the bag  appears, let us consider  a temperature 
independent surface tension:  $a_s (T) = a_{\rm o} > 0.$  
For this case,  Eq.  (\ref{Neleven}) shows that the  bag temperature  is decreasing function of its volume, 
whereas, according to  Eq.  (\ref{Nten}),   the energy of the bag is an increasing function
of the bag volume.  Therefore,  the bag's heat capacity, defined as ${\partial E}/{\partial T}$,
is negative.  This is shown schematically in Fig.~3.

For a  formal analysis of the heat capacity of the bag  it is necessary to use  Eqs. (\ref{Nseven}) and (\ref{Nnine}). From these equations one  can find the heat capacity of the bag at  
constant pressure $C_p$ and at constant volume $C_V$ as:
\begin{eqnarray}\label{Ntwelve}
\hspace*{-0.8cm} 
&& C_p   \equiv   T\hspace*{-0.1cm} \left( \frac{ \partial  S}{  \partial T} \right)_p \hspace*{-0.2cm} = 
C_V ~ - ~ \frac{ 3\, T V^\frac{4}{3} }{ 2\,a_s(T) } 
\left[  4 \sigma  T^3 -    \frac{2}{ V^\frac{1}{3} } \,  \frac{d~  a_s }{d~T}   \right]^2 \hspace*{-0.15cm}, \hspace*{0.15cm}\\
\label{Nthirteen}
\hspace*{-0.8cm} 
&& C_V   \equiv  T\hspace*{-0.1cm} \left( \frac{ \partial  S}{  \partial T} \right)_V \hspace*{-0.25cm} =  
4 \sigma \,T^3 \, V ~ -  ~  T\,  V^\frac{2}{3} \, \frac{d^2~ a_s }{d~ T^2}  \,. 
\end{eqnarray}
In evaluating the expression for $C_p$ we used an explicit form of the derivative 
\begin{eqnarray} 
\left( \frac{ \partial  V}{  \partial T} \right)_p  & \equiv &
- \left( \frac{ \partial  p}{  \partial T} \right)_V 
 \left( \frac{ \partial  p}{  \partial V} \right)_T^{-1} =  \nonumber  \\
\label{Nfourteen} 
&  - &  \frac{ 3\, V^\frac{4}{3} }{2\, a_s(T) }  
\left[ 4 \sigma T^3 -    \frac{2}{ V^\frac{1}{3} } \,  \frac{d~  a_s }{d~T}   \right] \,.
\end{eqnarray}

From Eqs.  (\ref{Ntwelve}) and (\ref{Nthirteen})  it is clearly seen that for any $T$  where  
$ a_s(T) \ge 0$ there may exist  a range of parameters for which 
the heat capacity $C_p$,  corresponding  to the bag  equilibrium in  vacuum,  is negative.
This leads to  a  ``convex intruder ''  in the entropy or an unusual behavior of  its 
second derivative:
\begin{equation} \label{Nfifteen}
\left( \frac{ \partial^2  S}{  \partial E^2} \right)_{p =0}   = ~ - ~ \frac{1}{T^2 \, C_p} \,, 
\end{equation}
which becomes positive  for this range of parameters.

In the  literature on this subject  it is argued  
\cite{Gross:97,Moretto:2002gm,Moretto:2003ii} that all small systems 
(comparable in size with the range of the  prevailing force) should show this effect. 
However, we  stress that  a convex intruder
in the bag model  with surface tension exists not for small systems, but  for  large ones  and   does not disappear 
in thermodynamic limit.
This behavior can be verified by examining the decay products  of  heavy resonances:
heavier resonances should decay into light hadrons of a lower temperature  (but never lower than $T(R = \infty)$).

Let us now demonstrate the appearance of a convex intruder  in a few  simple  cases. 
First we  consider the case of constant surface tension,  i.e. $a_s (T) = a_{\rm o} > 0$, 
in more detail.
Substituting  $ a_{\rm o}  $ into  Eq.  (\ref{Nfourteen}),  one obtains that  
$\left( \frac{ \partial  V}{  \partial T} \right)_p < 0$.   Since the heat capacity at $p = 0$ is  defined as  
$ C_p   \equiv   \left( \frac{ \partial  E}{  \partial V} \right)_{p=0}    \left( \frac{ \partial  V}{  \partial T} \right)_{p=0}  $,  its  sign  is  opposite to the sign of the derivative 
$\left( \frac{ \partial  E}{  \partial V} \right)_{p=0} $, which can be found from the 
expression for the energy of the bag:
\begin{equation} \label{Nsixteen}
E \biggl|_{p=0}  =    4\, B V~ +~  3\, a_{\rm o}  V^{ \frac{2}{3}} \quad \Rightarrow \quad 
\left( \frac{ \partial  E}{  \partial V} \right)_{p=0}  =  4\, B~ + ~  \frac{2\, a_{\rm o} }{V^{ \frac{1}{3}}  } ~ > ~0   \,.
\end{equation}
Thus,  in the case of a constant surface tension the heat capacity at  $p = 0$ is negative which 
corresponds to  a  convex intruder.

\begin{figure}

  \begin{minipage}[t]{16.cm}
   \vspace{0.0cm}
  \hspace*{-0.25cm}   \includegraphics[width=16.6cm,height=15.0cm]{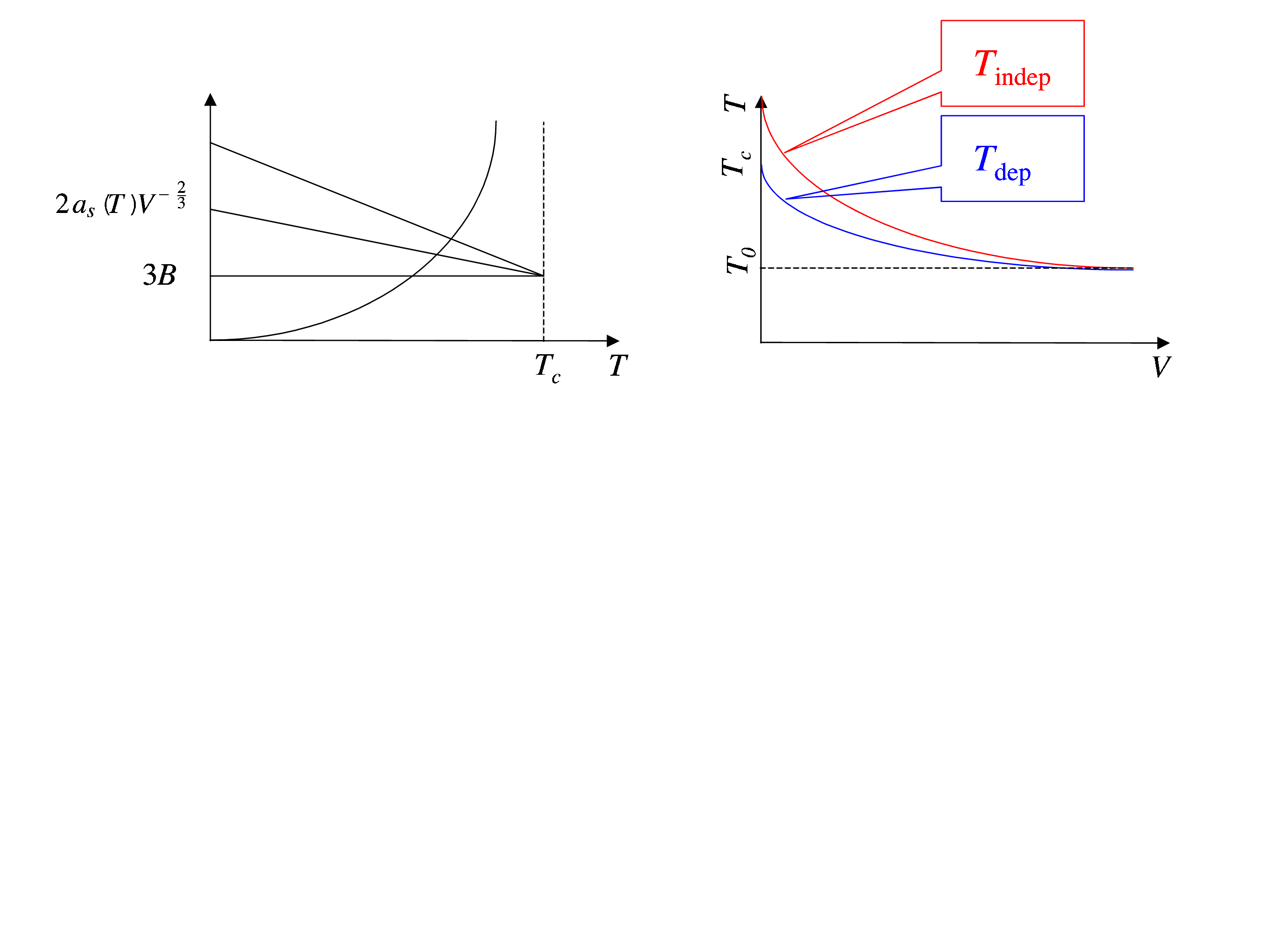}
   \vspace*{-9.0cm}
   
 \noindent  
{{\bf FIGURE 4.~ Left panel:} The bag temperature $T_H$ as a graphical solution of  Eq.  (\ref{Nseventeen}) 
for the linear $T$ dependence of the bag surface tension.  The left  hand side of 
Eq.  (\ref{Nseventeen})  is  shown by  a  bi-quadratic  parabola $ \sigma T_H^4$ and 
its  right hand side is depicted by the straight lines for different values of the bag volume $V$. 
The solution of  Eq.  (\ref{Nseventeen})  is found as an intersection point between the 
parabola and the straight line. \\
{\bf Right panel:}  Shows schematically the range of available temperatures of the bag for  $T$ independent  (red curve) and  for the linear $T$ dependent (blue curve) surface tension of the bag. 
See text for the details. 
}
 \end{minipage}
\label{fig4}
\end{figure}

Now we consider  a  surface tension with a  linear  $T$ dependence  in a spirit of the 
Fisher droplet model  \cite{Fisher:67} 
or using a more elaborate approach  of  the recently solved  
``Hills and Dales Model''  for  surface deformations \cite{Bugaev:04b, Bugaev:05a}:
$a_s (T) =  c_{\rm  o} \frac{(T_c - T)}{T_c}$, which is defined for
the  temperatures  not  above  the critical 
temperature $T_c$.
 
Introducing the notation $B \equiv   \frac{\sigma}{3}  T_{\rm o}^4$, one can rewrite 
the equilibrium  condition of the bag
$p ( T_H)  = 0$  as follows:
\begin{equation} \label{Nseventeen}
\sigma   T_H^4  = \sigma T_{\rm o}^4 ~+~ 2\,  c_{\rm  o} \frac{T_c - T_H}{T_c  \, V^{ \frac{1}{3} }   } 
\,,
%
\end{equation}
which should be solved for the bag  temperature $T_H (V)$. 

For  positive values of $T_H$ the left hand side of Eq. (\ref{Nseventeen}) is a monotonically 
increasing function of  $T_H$, whereas its right hand side is a monotonically decreasing function 
of $T_H$ (see the left panel of Fig.~4). Therefore, there can exist a single intersection point of these two functions for any 
positive value of  bag volume $V$.  Using Eq.   (\ref{Nseventeen}),  one can show that the inequality  $T_H \le T_c$ is 
always fulfilled,  if   $T_c  > T_{\rm o}$. 
Moreover, one can also show that the allowed interval of the bag temperatures is between 
$T_{\rm o}$ and  $T_c$  with limiting cases  $T_H (V \rightarrow 0) \rightarrow T_c$ 
and $ T_H (V \rightarrow \infty) \rightarrow T_{\rm o}$ (see the right  panel of Fig.~4). 
Similarly   from  Eq.  (\ref{Nfourteen}) one finds  that the bag temperature decreases,
while bag volume  grows, i.e. 
$\left( \frac{ \partial  V}{  \partial T} \right)_p < 0$ for any $V$ and  any  $T_H \le T_c$.
Since the range of allowed bag temperatures is bound between $T_{\rm o}$ and   $T_c $,
then from Eq. (\ref{Ntwelve}) one can immediately see that  for any $T_H \le T_c$ 
the  heat capacity of the bag  at  $p = 0$ is negative  for  large  volumes. 
Thus,  in the  case of a  linear $T$ dependence of  the surface tension of the bag  
the  convex intruder  exists for large volumes of the bag. 
In fact,  this  proves  the following statement:
{\it  if  the surface tension $a_s (T)  \ge 0$ is 
a regular function of $T$ that  $ \frac{d~  a_s }{d~T}  \le 0$ and $ \left| \frac{d^2~  a_s }{d~T^2} \right| $
is finite provided that  the solution $T (V)$ of  Eq.  (\ref{Neleven})  does not vanish in
the  limit $V \rightarrow \infty$, then in this limit  the heat capacity at constant $p = 0$
is negative and  $sign~C_p  = sign \left( \frac{ \partial  V}{  \partial T} \right)_p < 0$. 
}


\vspace*{-0.3cm}

\section{Stability of a Gas of Bags}

\vspace*{-0.2cm}

A gas of resonances (bags) is frequently considered  either in equilibrium or in 
 transport   problems. 
In our previous papers \cite{Moretto:05,Bugaev:05}  (see also  preceding  sections)  we have shown that 
an ordinary bag (no surface energy) is nearly indifferent to fragmentation into 
smaller bags. In fact, under rather general conditions it appears that there  is 
a  mild  tendency for a gas of bags to collapse into a single one.
We show now that the introduction of the surface leads to an even stranger tendency for a gas of bags toward collapse. 

Let us assume an arbitrary mass distribution in a gas of bags, and for simplicity, let us assume 
that the gas is confined in a fixed volume with its decay products (say pions). 
The gas cannot be isothermal since the smaller bags have larger temperature than the big ones. 
Thus the smaller bags evaporate  first and  their evaporation products are absorbed by the
larger bags until only one remains.  It may be argued that isothermicity  can be achieved by 
having all the bags to be of the same size.  But this situation is clearly unstable. Any small
perturbation in size will lead to a catastrophic collapse of all bags into a single one. 

\vspace*{-0.3cm}

\section{Decay  of a Bag}

\vspace*{-0.2cm}

A hot  bag,  unless   constrained by conserved quantities, must decay. 
As it decays, the instantaneous spectrum of the decay products  indicates  the bag's
instantaneous temperature.  Without surface effects the bag temperature is constant and the overall 
spectrum and the instantaneous spectrum  is  the same. 

With the surface effects, as the bag decays and becomes smaller, its temperature increases. 
Therefore the overall spectrum integrated over the overall decay must differ from the instantaneous
spectrum associated with each temperature. 
The  shape  deviation of the overall spectrum from that of an instantaneous spectrum at fixed temperature  may be an interesting observable  to  characterize both the effect and the 
magnitude  of  the surface energy and its  temperature dependence.
It is amusing  to notice the similarities  
with  a black hole  and its temperature  as it decays 
through the Hawking radiation.



\vspace*{-0.3cm}

\section{Conclusions}

\vspace*{-0.3cm}

In  Refs. \cite{Moretto:05, Bugaev:05}  we generalized the  SBM  results \cite{Frautschi:71} 
to  systems  of finite energy by showing  explicitly
that  even  a single resonance with  the Hagedorn mass spectrum degeneracy,
i.e. {\it a Hagedorn thermostat,}  keeps an almost constant   
temperature  close to $T_{\cal H}$  for  any  number of Boltzmann particles $3 < N_B  \le N_B^{kin} $. 
For the  high energy limit  $E \gg m_{\rm o}$ this means that 
a single  Hagedorn resonance defines 
the temperature of the system to be only slightly  different from $T_{\cal H}$ until 
 the energy of the Hagedorn thermostat is almost negligible compared to $E$.
In contrast to the grand canonical formulation of the original SBM  \cite{Frautschi:71},
in the presence of a Hagedorn thermostat 
the  temperature $T_{\cal H}$ 
can be reached  at  any energy density.

The thermostatic  nature of a Hagedorn system  
obviously
 explains the ubiquity of both the inverse slopes of  measured transverse mass spectra 
\cite{Gazdzicki:04} and hadronization temperature found in  numerical simulations 
of hadrons
created in elementary particle collisions at high  energies \cite{Becattini:Can,Becattini:1, Becattini:2}.
 By a direct evaluation of the microcanonical partition
we showed that  in the presence of a single Hagedorn thermostat  the energy spectra of particles become 
exponential 
 with no  additional assumptions, e.g. {\it  phase space dominance} \cite{PSDominance} or
{\it string tension fluctuations} \cite{Strings}.
Also the limiting temperature found in the URQMD calculations made in a  finite box 
\cite{URQMD:Box} can be explained by the effect of the Hagedorn thermostat. 
We expect that, if the string parametrization of the URQMD in a  box \cite{URQMD:Box} was done microcanonically instead of grand canonically,   the same behavior  would be found.

The Hagedorn thermostat model
generalizes the statistical hadronization model which successfully
describes the particle multiplicities in nucleus-nucleus and elementary collisions \cite{Becattini:Can,Becattini:1, Becattini:2}.
The statistical hadronization model  accounts for   the decay of heavy resonances 
(clusters in terms of Refs. \cite{Becattini:Can,Becattini:1, Becattini:2}) only  and does not consider the additional
particles, e.g. light hadrons, free quarks and gluons, or other heavy resonances.
As we showed,
the splitting of a single
heavy Hagedorn resonance into several  does not
change the temperature of the system. 
This finding  justifies the main assumption of 
the canonical formulation of the statistical hadronization model \cite{Becattini:Can} that smaller clusters   
may be reduced 
to a single large cluster. 
 Also our approach naturally explains why a sophisticated transport model \cite{Pawel:05},
which treats the hadronic reactions microscopically, leads to the thermal equilibration
at the Hagedorn  temperature $T_{\cal H}$ and to a chemical composition of hadrons given 
by the equilibrium values of particle concentrations. 
Thus,  
recalling the  MIT Bag model interpretation of 
the Hagedorn mass spectrum  \cite{Kapusta:81, Kapusta:82}, we conclude that 
 quark gluon matter confined in  heavy resonances (hadronic bags)  
controls the  temperature of surrounding particles close to $T_{\cal H}$, and, therefore,
this temperature can  be considered as 
a coexistence  temperature for  confined color matter and hadrons. 
Moreover, as we showed, the emergence  of a coexistence temperature does
not require the actual deconfinement of the color degrees of freedom,
which, in terms of the Gas of Bags Model  \cite{Gorenstein:81}, is equivalent to 
the formation of the infinitely large
and infinitely heavy hadronic bag.

Within the framework of  the Hagedorn thermostat model we found that even for a single Hagedorn thermostat and 
 $a >  \frac{3}{2}  $
the system's temperature $T = T^*(m_H^+)$  as a function of $N_B$ remains almost  constant    for 
 $N_B < N_B^{kin}$, reaches a maximum  at  $N_B^{kin} $  and rapidly decreases  
 for  $N_B > N_B^{kin} $ (see Fig. 1). 
For $a <  \frac{3}{2} $ the temperature  has a plateau $T =  T^*(m_H^+)$ for $N_B <  N_B^{kin}$,
and rapidly    decreases  for 
$N_B > N_B^{kin} $.
If such characteristic behavior 
of the hadronization temperature or the hadronic inverse slopes 
can be measured as a function of event multiplicity, it may be possible to experimentally
determine  the value of  $a$. 
For  quantitative predictions
it is necessary to include more hadronic species into the model, but this will not change 
our  result.

If we apply the Hagedorn thermostat model  to elementary particle collisions at high energy, then, as  shown above, 
the temperature of  created particles will  be  defined by the
most probable mass of the Hagedorn thermostat. 
If the most probable resonance mass 
grows  with  the energy of collision,  then the hadronization  temperature should decrease (increase)
to $T_{\cal H}$ for $a >  \frac{3}{2}  $ $(a <  \frac{3}{2}  )$. 
Such a decrease is  observed   in 
reactions of elementary particles  at  high energies, see Table 1 in Ref.
\cite{Becattini:2}.

Further we discussed the effects  of the surface energy on the properties of   a bag (heavy 
resonance) in  vacuum.
We showed that in the presence of  non-zero surface tension the temperature of the bag
(and  any system in thermal contact it) acquires a volume dependence, so that smaller bags 
are hotter. The temperature of large bags cannot be smaller than the Hagedron temperature.  
Under not too restrictive conditions we found that  the heat capacity  of large bags
at zero pressure 
is negative, i.e.  such bags have abnormal behavior of the second derivative of  entropy 
with respect to  energy. 
These unusual properties lead to an instability of  any number of bags other than  one. 
We argued that the temperature of the decay products of  the  evaporating bag should grow during
the evaporation process, which, hopefully, can be observed.

In order to apply 
these results in a more physical fashion to the quark gluon plasma formation in relativistic 
nucleus-nucleus collisions (where
the excluded volume effects  are known to be important  
\cite{Vol:1, Gorenstein:81, SPS:Chem, SPS:Chem1, SPS:Chem2,RHIC:Chem}  
for all hadrons) 
 the eigen volumes of all particles  should be incorporated into the model.
For pions this should be done in relativistic fashion \cite{Bugaev:00a}.
Also the effect of finite  width of Hagedorn resonances may be important \cite{SBM:width}  and should  be studied.

\end{document}